# ARC Sort: Enhanced and Time Efficient Sorting Algorithm

**Ankit R. Chadha**
Electronics & Telecommunication Engineering
Vidyalankar Institute of Technology
Mumbai, India

**Rishikesh Misal**
Department of Computer Engineering.
Vidyalankar Institute of Technology
Mumbai, India

**Tanaya Mokashi**
Department of Computer Engineering.
Vidyalankar Institute of Technology
Mumbai, India

**Aman Chadha**
Electrical and Computer Engineering,
University of Wisconsin Madison, WI, USA

## ABSTRACT

This paper discusses about a sorting algorithm which uses the concept of buckets where each bucket represents a certain number of digits. A two dimensional data structure is used where one dimension represents buckets i.e; number of digits and each bucket's corresponding dimensions represents the input numbers that belong to that bucket. Each bucket is then individually sorted. Since every preceding bucket elements will always be smaller than the succeeding buckets no comparison between them is required. By doing this we can significantly reduced the time complexity of any sorting algorithm used to sort the given set of inputs.

## Keywords
Sorting, Bucket, Enhanced Selection, ARC Sort

## 1. INTRODUCTION
One of the most recurrent problems in the field of computer science is sorting a list of items, numbers or strings. It refers to the arranging of numerical or alphabetical or character data in statistical order (either in increasing order or decreasing order) or in lexicographical order (alphabetical value like addressee key) [1-3].Computer Applications routinely use sorting before implementing search, merge or normalize functions.The most commonly used algorithms like Bubble Sort [4], Selection Sort & Insertion Sort are time consuming unlike the enhanced versions which are considered as complex algorithms. There is a positive correlation between complexity and effectiveness of an algorithm [5]. Hence, much work is being done to improve the efficiency of algorithms and allocation of resources, as execution of comparisons, swaps and assignment operations are involved in sorting algorithms.

Complexity of programming, computation & storage besides presence of pre-sorted data affect the choice of sorting algorithm to use.

Our purpose behind using ARC sort is to reduce the number of iterations required to sort the given number of elements. Reduce the number of unnecessary comparisons and swaps which will optimize any sorting algorithm. We divide the array of input numbers by placing them in a bucket. This bucket corresponds to the number of digits present in the input element. This will reduce the number of unnecessary comparisons by a fair bit and will certainly reduce the number of swaps.

## 2. ENHANCED SELECTION SORT
In enhanced selection sort, first the maximum element amongst all the other elements of the array is placed in its right position. In every iteration after finding the maximum element the size of the array is reduced by 1. Enhanced selection sort is better than selection sort because it reduces the number of swaps needed to sort the elements, this makes the algorithm more stable. The time complexity remains the same as that of selection sort that is $O(n^2)$[7].

Procedure for enhanced selection sort

1. Call enhanced selection sort function with the input array and its size
2. Find the maximum element and place it in its correct position
3. Decrement the size by 1
4. Call enhanced selection sort function recursively with the reduced size

## 3. CONCEPT OF ARC SORT
Consider a two dimensional array A[][] with k number of columns and n number of rows
   Where,
      k - pre-estimated highest number of digits amongst the given input numbers
      n - number of input numbers
Another array index[] of length k which indicates that how many numbers are present in each bucket of array A[][]
Consider a value of k which represents the highest number of digits
For each input number calculate the number of digits. Now store that number at the bucket corresponding to the number of digits. Increment the array index[] whose position is represented by number of digits. Repeat this process for each input number and populate each input number in its appropriate position. If the input array contains negative numbers dedicate one bucket for all the non-positive numbers. Now since negative numbers are also present, the first bucket is reserved for those numbers that is the $0^{th}$ bucket. In general, if a number has m number of digits it will be placed in $m^{th}$ bucket of array A[][]
Once all the input numbers are placed in buckets, now sort each bucket individually using Enhanced Selection sort. The purpose of dividing each element into buckets of number of digits is it reduces the number of passes needed to sort an array.

The other purpose of diving the array of given input numbers is that every i<sup>th</sup> bucket numbers will always be less than all the numbers present in j<sup>th</sup> bucket where i<j that is every preceding bucket elements will always be smaller than all its succeeding buckets. This reduces the sorting task by a huge margin. If the input numbers are of different ranges dividing the input array will help increase the efficiency of the sorting algorithm used.

## 4. FLOW CHART

Below is the flow chart for the ARC algorithm
Terms used,
Main() – represents code present in main function of the program
Label A – call to count(int) function which computes the number of digits present in the given integer
Label B – return back to calling function i.e main()
Label C – call to sort(int [][],int) function to sort the given array
Label D – return back to calling function i.e main()

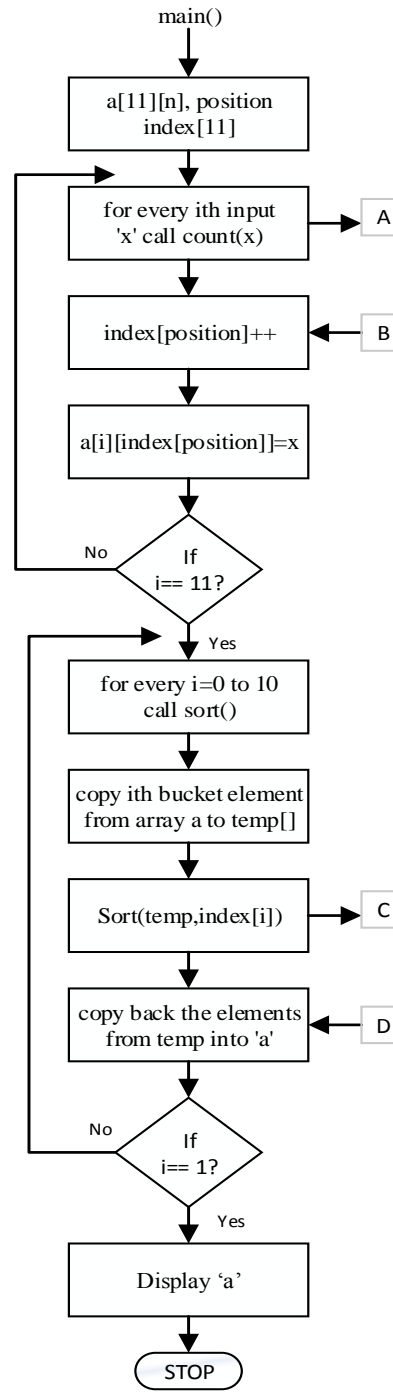

32

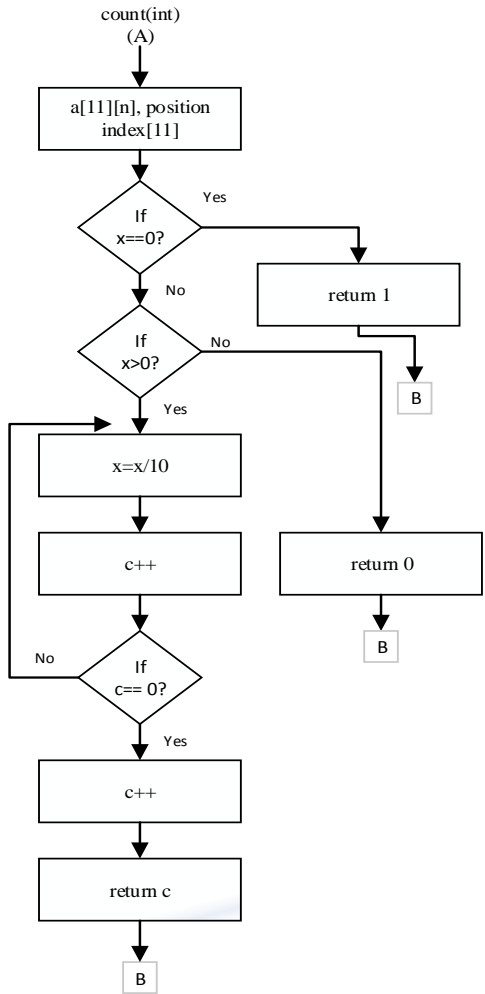

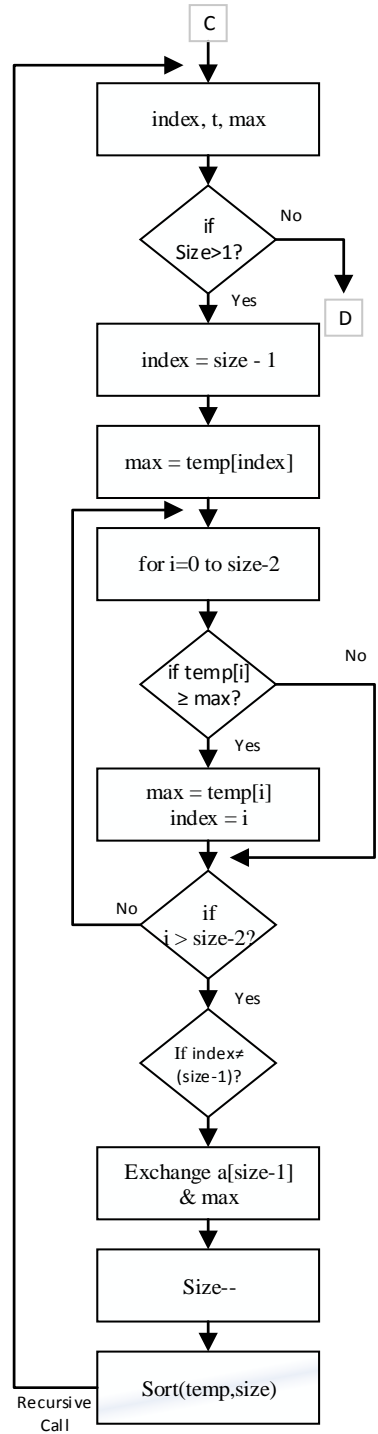

## 5. Pseudo Code

Terms: x – current input element , k – number of digits of max number, index[k] – every element initialized to -1, a[k][n]

   *main()*

   1. *For i=0 to n-1*
   2. *Position = count(x) // count is function which will calculate the number of digits of x*
   3. *Index[position]++;*

4. *a[position][index[position]]=x*
5. *end for*
6. *for i=0 to k*
7. *if(index[i]>0)*
8. *sort(a[i][],index[i])*
9. *end if*
10. *end for*

*sort(a[][],size)*

1. *if(size>1)*
2. *index=size-1;*
3. *max=a[index]*
4. *for i=0 to size-2*
5. *if(a[i]>=max)*
6. *max=a[i]*
7. *index=i*
8. *end if*
9. *end for*
10. *if(index!=size-1)*
11. *temp=a[size-1]*
12. *a[size-1]=max*
13. *a[index]=temp*
14. *end if*
15. *size =size -1*
16. *sort(a,size)*
17. *end if*

## 6. Example

Input:

| 349 | 34 | -72 | 22 | 14 | -1 |
|---|---|---|---|---|---|
| 0 | 1 | 2 | 3 | 4 | 5 |

Consider
K = 5 = highest number of digits present in the data set.
M = 6 = no. of elements
A[5][6]
Index[6]
For E(i=0 to 5)
Index[i] = -1

1) X = a[0] //input element when i=0
X = 349
position = count(X), //count() will calculate the numbers of digits present in the number
position = 3 // since x=349 contains 3 digits
index[position]++ // increment the count for the corresponding bucket in this case bucket 3
a[position][index][position] = X, //place the input number in its appropriate position

index[ ]

| -1 | -1 | -1 | 0 | -1 | -1 |
|---|---|---|---|---|---|
| 0 | 1 | 2 | 3 | 4 | 5 |

| a | 0 | 1 | 2 | 3 | 4 |
|---|---|---|---|---|---|
| 0 |   |   |   | 349 |   |
| 1 |   |   |   |   |   |
| 2 |   |   |   |   |   |
| 3 |   |   |   |   |   |
| 4 |   |   |   |   |   |
| 5 |   |   |   |   |   |
| 6 |   |   |   |   |   |

2) X = a[1] // input element
X = 34
position = count (X);
position = 2;
index[position] ++;
a[position][index][position] = X;

index[ ]

| -1 | -1 | 0 | 0 | -1 | -1 |
|---|---|---|---|---|---|
| 0 | 1 | 2 | 3 | 4 | 5 |

| a | 0 | 1 | 2 | 3 | 4 |
|---|---|---|---|---|---|
| 0 |   |   | 34 | 349 |   |
| 1 |   |   |   |   |   |
| 2 |   |   |   |   |   |
| 3 |   |   |   |   |   |
| 4 |   |   |   |   |   |
| 5 |   |   |   |   |   |
| 6 |   |   |   |   |   |

3) X = a[2]
X = -72
position = count (X);
position = 0; //since the input number is negative its bucket is reversed that is the 0[th] bucket
index[position] ++;
a[position][index][position] = X;

index[ ]

| 0 | -1 | 0 | 0 | -1 | -1 |
|---|---|---|---|---|---|
| 0 | 1 | 2 | 3 | 4 | 5 |

| a | 0 | 1 | 2 | 3 | 4 |
|---|---|---|---|---|---|
| 0 | -72 |   | 34 | 349 |   |
| 1 |   |   |   |   |   |
| 2 |   |   |   |   |   |
| 3 |   |   |   |   |   |
| 4 |   |   |   |   |   |
| 5 |   |   |   |   |   |
| 6 |   |   |   |   |   |

4) X = a[3]
X = 22
position = count (X);
position = 2;
index[position] ++;
a[position][index][position] = X;

index[ ]

| 0 | -1 | 1 | 0 | -1 | -1 |
|---|---|---|---|---|---|
| 0 | 1 | 2 | 3 | 4 | 5 |

| a | 0   | 1 | 2  | 3   | 4 |
|---|-----|---|----|-----|---|
| 0 | -72 |   | 34 | 349 |   |
| 1 |     |   | 22 |     |   |
| 2 |     |   |    |     |   |
| 3 |     |   |    |     |   |
| 4 |     |   |    |     |   |
| 5 |     |   |    |     |   |
| 6 |     |   |    |     |   |

5) X = a[4]
X = 14
position = count (X);
position = 2;
index[position] ++;
a[position][index][position] = X;

index[ ]

| 0 | -1 | 2 | 0 | -1 | -1 |
|---|----|---|---|----|----|
| 0 | 1  | 2 | 3 | 4  | 5  |

| a | 0   | 1 | 2  | 3   | 4 |
|---|-----|---|----|-----|---|
| 0 | -72 |   | 34 | 349 |   |
| 1 |     |   | 22 |     |   |
| 2 |     |   | 14 |     |   |
| 3 |     |   |    |     |   |
| 4 |     |   |    |     |   |
| 5 |     |   |    |     |   |
| 6 |     |   |    |     |   |

6) X = a[4]
X = 14
position = count (X);
position = 2;
index[position] ++;
a[position][index][position] = X;

index [ ]

| 0 | -1 | 2 | 0 | -1 | -1 |
|---|----|---|---|----|----|
| 0 | 1  | 2 | 3 | 4  | 5  |

| a | 0   | 1 | 2  | 3   | 4 |
|---|-----|---|----|-----|---|
| 0 | -72 |   | 34 | 349 |   |
| 1 |     |   | 22 |     |   |
| 2 |     |   | 14 |     |   |
| 3 |     |   |    |     |   |
| 4 |     |   |    |     |   |
| 5 |     |   |    |     |   |
| 6 |     |   |    |     |   |

7) X = a[5]
X = -1
position = count (X);
position = 0;
index[position] ++;
a[position][index][position] = X;
index[ ]

| 1 | -1 | 2 | 0 | -1 | -1 |
|---|----|---|---|----|----|
| 0 | 1  | 2 | 3 | 4  | 5  |

| a | 0   | 1 | 2  | 3   | 4 |
|---|-----|---|----|-----|---|
| 0 | -72 |   | 34 | 349 |   |
| 1 | -1  |   | 22 |     |   |
| 2 |     |   | 14 |     |   |
| 3 |     |   |    |     |   |
| 4 |     |   |    |     |   |
| 5 |     |   |    |     |   |
| 6 |     |   |    |     |   |

Sort ( )
i) index[0] > 0 // since the number of elements present in $0^{th}$ bucket are greater than 0 the elements need to be sorted
sort (a[0][ ],index[0])  // call function sort with the $0^{th}$ bucket elements are the size indicated by the index array

Output:

| a | 0   | 1 | 2  | 3   | 4 |
|---|-----|---|----|-----|---|
| 0 | -72 |   | 34 | 349 |   |
| 1 | -1  |   | 22 |     |   |
| 2 |     |   | 14 |     |   |
| 3 |     |   |    |     |   |
| 4 |     |   |    |     |   |
| 5 |     |   |    |     |   |
| 6 |     |   |    |     |   |

ii) index [1] = =-1 // since no elements are present in the $1^{st}$ bucket sort function will not be called
skip

iii) index [2] > 0
sort (a[2][ ], index[2])
output:-

| a | 0   | 1 | 2  | 3   | 4 |
|---|-----|---|----|-----|---|
| 0 | -72 |   | 14 | 349 |   |
| 1 | -1  |   | 22 |     |   |
| 2 |     |   | 34 |     |   |
| 3 |     |   |    |     |   |
| 4 |     |   |    |     |   |
| 5 |     |   |    |     |   |
| 6 |     |   |    |     |   |

iv) index [3]  = 0  // since only 1 element is present in the $3^{rd}$ bucket sorting is not required therefore skip this iteration
skip
v) index [4] = = -1
skip

Display Array "a" bucket wise

| -72 | -1 | 14 | 22 | 34 | 349 |
|-----|----|----|----|----|-----|
| 0   | 1  | 2  | 3  | 4  | 5   |

## 7. Performance Analysis
### 7.1 Best Case
If k=n , where k- number of digits of maximum element in the data set & n- number of data elements in the data set
and If every bucket has exactly one element , the time complexity of the ARC algorithm is,
$T(n) = O(n^2/k)$
$= O(n^2/n)$
$= O(n)$

### 7.2 Average Case
If there are k buckets and each bucket contains equal number of elements that is,
Each bucket contains n/k elements
Time complexity of ARC algorithm becomes,

$T(n) = O(n^2/k^2)$

### 7.3 Worst Case

If there are k buckets present but all the elements in the data set belong to the one particular bucket
Therefore the time complexity of ARC algorithm becomes same as that of Modified Selection Sort
$T(n) = O(n^2)$

**Table 1: Comparison with other Sorting algorithm**

|  | ARC Sort | Insertion Sort | Selection Sort | Bubble Sort |
|---|---|---|---|---|
| Best Case | $O(n)$ | $O(n)$ | $O(n^2)$ | $O(n^2)$ |
| Average Case | $O(n^2/k)$ | $O(n^2)$ | $O(n^2)$ | $O(n^2)$ |
| Worst Case | $O(n^2)$ | $O(n^2)$ | $O(n^2)$ | $O(n^2)$ |

Where, k- number of buckets

**Table 2: Execution Time Comparison**

| No. of Elements | Elapsed Time (ms) | | | |
|---|---|---|---|---|
|  | ARC Sort | Selection Sort | Insertion Sort | Bubble Sort |
| 1,000 | 10 | 18 | 17 | 15 |
| 5,000 | 13 | 59 | 26 | 65 |
| 10,000 | 31 | 207 | 53 | 65 |
| 20,000 | 126 | 743 | 141 | 841 |

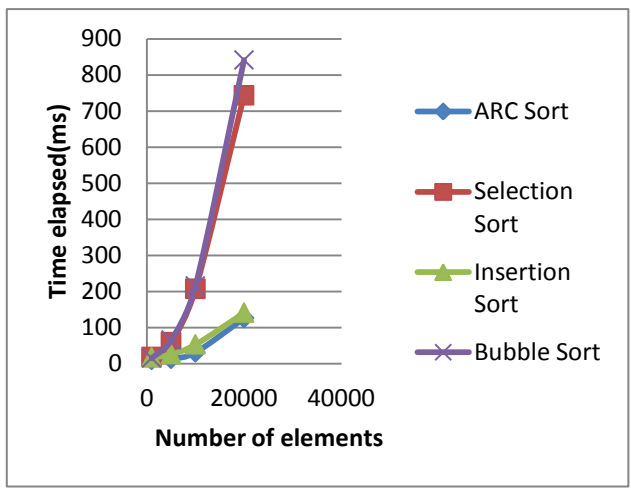

**Fig 1 : Elapsed time comparison for various sorting algorithms**

The above table & figure represents the comparison of ARC Sort with Selection, Insertion and Bubble Sort. The left hand side column is the number of elements required to be sorted. For each number of elements the given sorting algorithms are run. The corresponding time taken by each algorithm is as shown above. The input set used to sort was taken randomly & the average value of each run was taken as the actual time of execution.

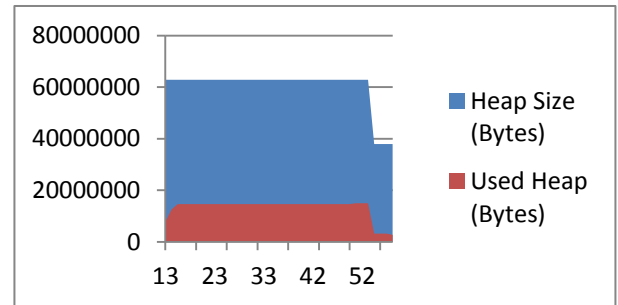

**Fig 2 : Memory Heap**

The above graph shows Memory heap used and allocated. The maximum heap allocated was greater than 60MB and the maximum amount of heap used is greater than 10MB but less than 15MB. The y-axis of the graph shows the amount of heap used, whereas the x-axis of the graph shows the amount of time elapsed.

### 8. CONCLUSION

The paper presents the concept & implementation of ARC Sort which not only reduces the running time as compared to Selection Sort, Insertion Sort and Bubble Sorting but is also efficient in performance as it reduces the number of unnecessary comparison and swaps by making use of 2 Dimensional Arrays.